# Trench Gate Power MOSFET: Recent Advances and Innovations

Raghvendra Sahai Saxena and M. Jagadesh Kumar

*Chapter 1*

# Trench Gate Power MOSFET: Recent Advances and Innovations

*Raghvendra Sahai Saxena and M. Jagadesh Kumar*

The trench gate MOSFET has established itself as the most suitable power device for low to medium voltage power applications by offering the lowest possible ON resistance among all MOS devices. The evolution of the trench gate power MOSFET has been discussed in this chapter, starting right from its beginnings to the recent trends. The innovations in the structural improvements to meet the requirements for an efficient operation, the progress in the fabrication process technology, the characterization methods and various reliability issues have been emphasized.

I. **INTRODUCTION**

From the very beginning in the history of mankind, there have been continuous and systematic scientific efforts to make our lives easy and comfortable. The development of power electronic devices is one of the steps towards that goal as these devices provide the means to control heavy electro-mechanical loads, industrial machines etc. to perform difficult tasks that would have been impossible otherwise. In these applications, a huge amount of electrical energy is required, which is acquired from an electrical source and stored in various elements and finally released in a controlled manner for our intended use of driving various devices and systems. The basic goal of controlling the flow of energy from its source to the load in an efficient manner with high reliability is accomplished by using the design concepts of electrical engineering with sophisticated analytical tools and efficient power electronic devices of small size, light weight and most importantly the low cost.

We have entered the age of limited energy resources amidst an emerging energy crisis. Therefore, highly efficient, rugged and reliable power electronic devices are required so that without consuming much energy, a variety of our needs may be fulfilled within the economical and manufacturing constraints. To meet the energy efficiency demand, there is a need for efficient power MOSFETs, capable of delivering high power without consuming a significant part of it. Out of a variety of possible options, trench gate architecture of power MOSFET provides the most efficient performance as a discrete power device for relatively low voltage applications due to their lowest achievable ON-state resistance (RON) than any other structure of otherwise similar specifications.

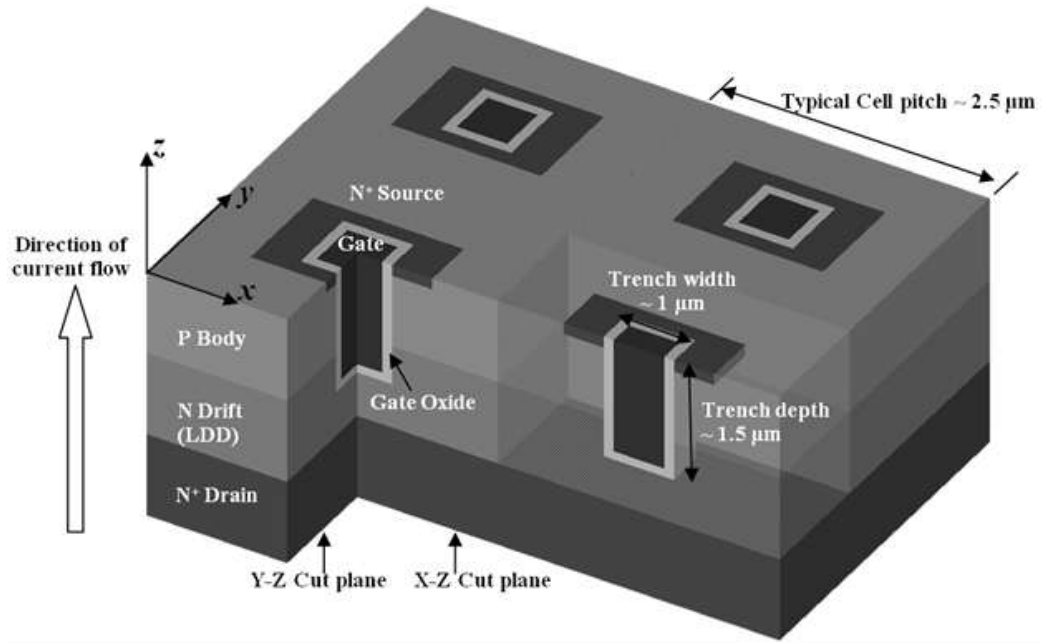

Fig. 1: Schematic 3D view of a trench gate power MOSFET showing typical dimensions.

A trench gate MOSFET is basically an attempt to make a complete chip conduct the current vertically from one surface to the other so as to achieve a high drive capability. It is realized by packing millions of trenches on a chip, deep enough to cross the oppositely doped 'body' region below the top surface. Each trench houses a gate dielectric and gate electrode to control the current conduction in its vicinity by the virtue of field effect. The schematic 3D view of a part of a trench gate MOSFET, with its cut sections along two vertical cut planes, is shown in Fig. 1 indicating four adjacent cells (each containing a trench). Similar to any other MOSFET, a trench MOSFET cell contains the drain, gate, source, body and the channel regions but exhibits a vertical direction of current flow. All the cells are connected to work in parallel in order to reduce the value of RON. In addition, it has a lightly doped drain (LDD) region between the channel and the drain to make it capable of sustaining large voltage in OFF-state condition. In the ON-state condition, the charge carriers are simply drifted through the LDD region towards the drain due to the potential difference across it. Therefore, the LDD region is also known as the 'drift region'.

The major design considerations of a trench gate power MOSFET are lowering of RON, enhancement of breakdown voltage (VBD), reduction in switching delays, enhancement in transconductance, enhancement in dV/dt capability, high damage immunity while switching large current in inductive loads and minimization of energy losses. In a given structure, all the parameters are technologically linked with each other and none of these can be adjusted independently. For example, thickness and doping concentrations of the drift and the body regions decide the VBD and RON. Similarly, the gate oxide thickness decides the threshold voltage (VTh), transconductance and gate capacitance. Therefore, the design of a trench gate MOSFET for a specific application is a tradeoff among its performance parameters. With the passage of time, various improvements have been suggested, analyzed and implemented to modify the basic device structure for improving these tradeoffs.

As far as energy efficiency is concerned, there are two types of energy losses in a power MOSFET that need to be suppressed. The first and foremost is the conduction loss arising from the non-zero RON and the other one is the switching loss caused by charging and discharging of the gate electrode while switching the device ON and OFF. This charge is known as gate charge (QG). These two losses are also interlinked. Any attempt to reduce the RON in a conventional structure results in an increased QG, and vise a versa. Therefore, a real improvement in device performance is considered only if the product of RON and QG is reduced. Therefore, RON.QG is called figure of merit (FOM).

## II. EVOLUTION OF TRENCH POWER MOSFETS: A BRIEF HISTORY

The semiconductor power devices have evolved in the second half of the 20th century [1-4], when the first significant device was demonstrated by Hall in 1952 using germanium mesa alloy junction [5]. This was the first step towards getting rid of the bulky, large and less reliable vacuum tube devices and therefore it attracted a lot of research interest in semiconductor electronic devices, resulting in a rapid development in the semiconductor technology. Soon after its first commercial launch in 1954 by Texas Instruments [6], Silicon based Bipolar Junction Transistor (BJT) has taken over the majority of the power device market by replacing the vacuum tube devices. However, later on it was realized that the BJT was also not suitable for various emerging power applications due to many problems associated with it [7, 8], e.g., its positive temperature coefficient of current that makes it susceptible to thermal runaway, the charge storage problem in the base that makes it a slow switch, secondary breakdown and the low input resistance that necessitates a large controlling current consumption. These shortcomings of BJTs motivated intense research for their improvements as well as search for a new and better device. In this direction, the metal oxide semiconductor (MOS) based insulated gate devices have been investigated and found to be the most promising from various perspectives. A MOSFET, due to its insulated gate, puts a very little load to the input circuitry and being a majority carrier device, it does not suffer from the charge storage problem. Furthermore, the negative temperature coefficient of current not only makes MOSFETs better from the thermal runaway point of view but also makes them suitable for parallel conduction. With all these advantages, people started exploring the suitability of the MOSFETs in the power electronic applications. The only difficulty found in the MOSFET as a power device was its large ON-resistance that results in conduction losses. It was quite attractive to overcome this difficulty in an otherwise near ideal power device, i.e., MOSFET that had the potential to replace the bipolar transistors. The main structures that have been proposed in early 1970's to implement the insulated gate controlled devices were LDMOS, VDMOS, and VMOS [9].

Around early 1980's the power MOS transistors had started competing bipolar power devices in power-handling capability when the technological improvements made it possible to successfully realize very short active channels and incorporation of a lightly doped region between the channel and the drain for improving both ON and OFF state device performance [9]. This structure was double diffused MOSFET (DMOSFET). Later on the V-grooved MOSFET (VMOSFET) had been developed as a potential candidate to meet the upcoming power requirements with lower conduction losses, but they were suited only for low voltage applications [10-21]. Development of VMOSFET was the first attempt of making a vertical MOSFET and is considered to be the major breakthrough in the power electronics because their large scale integration capability and the absence of parasitic JFET made them able to provide the lowest ON-resistance compared to other possible structures. However, the VMOSFET had a problem of low breakdown voltage due to the crowding of electric field lines at its sharp bottom. This problem was solved partially by smoothing the V-shaped bottom that resulted in the UMOSFET structure. By 1985, the advancements in the etching technology has enabled opening of rectangular grooves that resulted in the fabrication of a new vertical power MOSFET structure called rectangular grooved MOSFET (RMOS), in which the vertical channels along the sidewalls of the rectangular grooves were formed by a reactive ion-beam etching (RIBE) technique [22]. This is how the UMOSFET structure has further changed into its improved successor, the trench gate MOSFET. This structure was another major milestone of the progress of low ON-resistance and high packing density MOSFET structures and was demonstrated experimentally to have lower ON-resistance per unit area than other VMOS and DMOS structures [23-29].

### III. STRUCTURAL ADVANCEMENTS AND INNOVATIONS

Although the ON-resistance offered by the trench gate MOSFET has been the lowest among various possible structures, the applications keep demanding further reduction in ON-resistance with sufficiently high breakdown voltage and reliability. To meet this ever persistent demand, various modifications in the basic trench gate structure with improved fabrication technology have been incorporated. Such important milestones are discussed in subsequent subsections.

*A. Trench Optimization*

The problems associated with the trench have motivated the development of improved and optimized trench structures. The major considerations at the initial phase of their development were the depth of trenches and the sharp corners. In 1986, D. Ueda et. al. have realized the deep trench gate MOSFET having three times smaller RON than that of the best achievable structure with a shallow trench and have shown that the use of deep trenches had two additional advantages, i.e., low spreading resistance due to the increased accumulation layer area and complete elimination of parasitic JFET, reducing the RON [30]. Later on, the deep trenches have been further investigated experimentally as well as theoretically and found to have superior characteristics [31-34]. The deep trenches were also found to have better immunity for the repetitive inductive switching [35]. Another initial technological limitation was related to the sharp corners at the trench bottom having a very strong corner effect, i.e., crowding of electric field lines that exist even with the low fixed oxide charge density ($Q_f$). It was investigated that the doping concentration of higher than $10^{17}$ cm$^{-3}$ was necessary for the case of $Q_f = 10^{11}$ cm$^{-2}$ to minimize the inversion at the corners that, on the other hand, can adversely affect the leakage currents along the trench [36]. The thicker gate oxide has also been tried as an approach to control the corner effect, but was found not very effective. Finally, corner rounding was found to be the only effective way to get rid of the corner effect. In its implementation, the area loss due to corner rounding is minimized by properly adjusting some of the parameters like substrate doping concentration, oxide thickness and corner curvature. To overcome the technological problems associated with the trench corner shaping, a novel fabrication process technique of pull-back has been suggested that uses hydrogen annealed trench surface [37-41]. Due to the virtue of silicon migration, it results in highly reliable thin gate oxide along with corner rounding [38].

      The deep trenches with round corners were found suitable for lowering RON and increasing the breakdown voltage. However, as they extend deep into the drift region, the gate capacitances are enhanced affecting the switching delays. In a power MOSFET, the switching delay is a very important performance parameter and is mainly governed by the gate capacitance, especially the gate-to-drain capacitance, $C_{GD}$ that acts as miller capacitance in usual circuit configurations. Therefore, $C_{GD}$ has to be minimized to make the ON and OFF transitions fast. The WMOSFET is an attractive design that reduces $C_{GD}$ by using higher gate oxide thickness at the trench bottom and is realized using LOCOS and sub atmospheric CVD (SACVD) processes [42-45]. The structure was named as WMOSFET due to the 'W' like shape of its trench. The WMOSFET has been demonstrated experimentally to have a significant reduction in gate-drain charge $Q_{GD}$, a low RON and a good production process margin [43-45].

*B.     Body/Channel Profile Engineering*

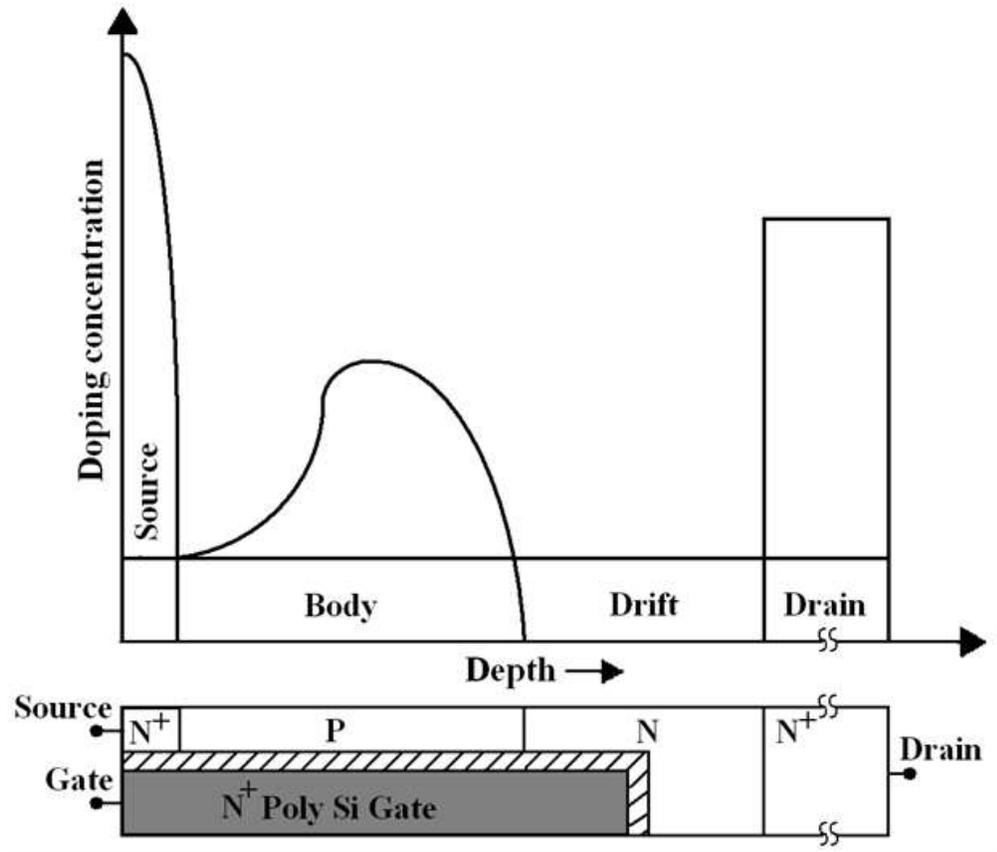

Fig. 2: Schematic doping profile of retrograde body trench MOSFET [46].

In order to achieve a low specific ON-resistance and low gate capacitance, a simple method of retrograde body profile has been suggested [46-48]. The retrograde body MOSFET (RBMOS) was realized by high energy implantation for body formation as shown in Fig. 2. In this structure, basically the doping concentration in the channel is adjusted in such a way that its drain-side gets highest concentration. This reduces the depletion width at the drain-side, resulting in the screening of drain voltage at the source-side and, therefore, allows a significant reduction in channel length without degrading the punch-through voltage. Low thermal budget and easy trench gate process are its additional benefits. Comparing with the conventional device, the specific ON-resistance and the figure-of-merit have been reported to improve by 38% and 70%, respectively, in the RBMOS device [46].

Another type of channel engineering that reduces the channel resistance is the use of hetero-epitaxy. SiGe epitaxial layer in the body region has been demonstrated to provide lower RON, but at the cost of a little degradation in thermal performance [49]. This degradation can be retrieved by incorporating carbon to realize the SiGeC body/channel [50]. A device with the SiGeC body shows significantly smaller temperature sensitivity of the drain current as compared to the Si and SiGe devices, while the higher drain current advantage of SiGe channel is still preserved in the SiGeC device.

### C. Drift Region Engineering

The drift region is designed to sustain a large drain to source voltage in OFF condition and, therefore, is doped lightly. However, due to the low doping concentration, its resistance becomes large resulting in higher ON-resistance of the device. The difficulty behind simultaneously meeting the requirement of low ON-resistance and high breakdown voltage becomes severe as these two parameters are found to be related by about 2.5th power relation [51-63]. Various novel structures like floating island, super junction and RESURF (reduced surface field) effect have been worked out to improve the trade-off between specific ON-resistance and breakdown voltage, such as oppositely doped floating island devices (FLIMOSFET and FITMOSFET) [51-59], super-junction/COOLMOS devices [60, 61], Opposite Doped Buried Regions (ODBR) MOSFET [62], and P-buried layer Schottky barrier diodes [63]. Incorporation of the floating islands inside the drift region of the device modifies the electric field distribution in the bulk in such a way that there exist several (depending on the number of buried islands) small peaks in electric field profile. This results in a significant lowering of the highest peak [56]. The formation of multiple peaks allows increasing the doping concentration up to the level at which the highest peak of electric field is just less than the breakdown field, thereby improving the relation between RON and VBD as well as reducing the FOM [59]. Additionally, the reverse recovery of the body diode becomes faster in these devices, reducing the switch OFF delays. On the other hand, there is a drawback of floating islands that the JFET effect is enhanced, resulting in the enhancement of RON under AC mode of operation. This issue was solved by the passive hole gates to control the minority carriers during AC operation that resulted in the RON equivalent under both AC and DC operations [59]. The trade-off between VBD and RON is also improved by using floating islands in elliptical forms [57-59].

Many efforts have been made for incorporating super junction or RESURF structure in trench gate MOSFETs, which is not straight forward in such non-planar devices. The first SJ trench structure was realized by T. Nitta et. al. by making the P and N stripe structure in alternating fashion fabricated by angled implantation in very deep trenches [64]. Though the structure was not actually the trench gated MOSFET, it was an important contribution as it confirmed the field relaxation possibility in vertical trench structures and motivated further research in this direction. Soon after this attempt, Y. Hattori et. al. have realized the first real SJ trench gate MOSFET [34], wherein an accurate control of the impurity concentration of N drift region was achieved by As- ion implantation at 10o angle in the deep trenches opened in a P-epilayer. The structure realized in [34] is shown schematically in Fig. 3. It gives 30% reduction in RON as compared with the conventional trench gate structures. A similar structure with slight modifications for thicker epilayers and proper care of the trench gate misalignments has been used to realize 200 V devices and showed 35% reduction in RON [65]. In another approach, instead of implanting N-type dopants, SJ trench structure was realized by multiple boron ion implantations in N-type drift region with energies up to 2 MeV, to obtain p-columns with almost flat sidewalls [66]. This has shown 30% reduction in the RON. A similar structure with split in the P column was realized with the similar approach, only by adjusting the implantation energies [67]. The continuous p-column body structure and the split body SJ structure are shown in Fig. 4(a) and (b) respectively. The split body device was reported to have better inductive switching immunity also. The SJ effect may also be realized by forming P-column (pillar) at the trench bottom. Such a SJ trench MOSFET has been realized by opening deep trenches and growing P-epilayer in the trench before making gate [68]. The structure has been analyzed further and it was shown that increasing the depth of the P-column may reducce RON by 70% [69]. A super 3D structure composed of the deep source N region, deep channel P region, the deep drift N region and the deep drain N region, arranged laterally in order, has been proposed [70-73] with the trenches deep enough to reach near the bottom of the drift region. This structure provides the advantage of wide channel, improving current drive capability. Another approach of deep trench filling with highly anisotropic epilayer growth has been demonstrated to realize a 200 V SJ MOSFET. The filling of high aspect ratio trenches was the major technological challenge in the realization of this structure, which was met by applying silicon and chlorine source gases simultaneously [74]. A further improvement in this structure has been suggested, wherein the trench gates were made orthogonal to the P/N columns, allowing independent control of trench gate pitch and p/n column pitch [75]. This structure has shown 36% lower figure of merit. Moreover, the unclamped inductive switching endurance of this SJMOSFET was found to be 30% larger than the conventional trench MOSFET.

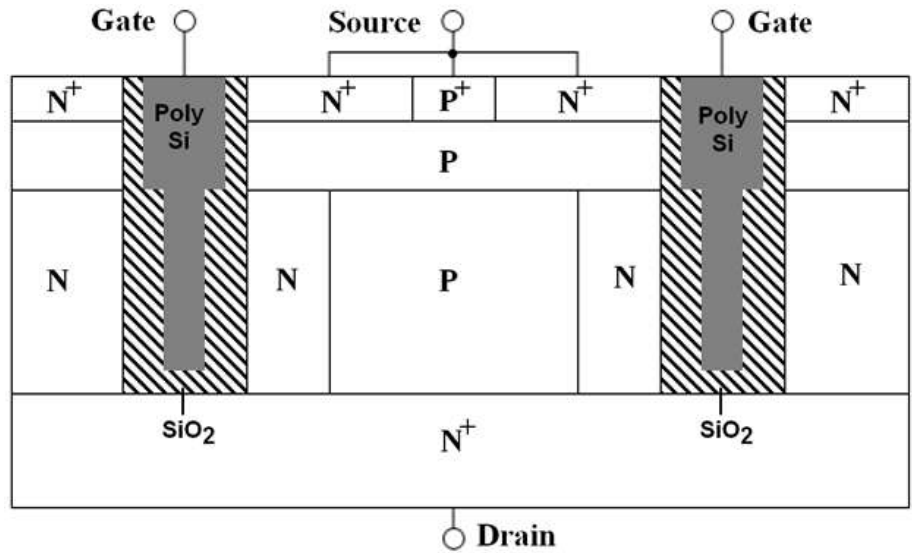

Fig. 3: A super junction trench gate power MOSFET realized using arsenic ion implantation in P-type epilayer at 10 degree angle in trenches [34].

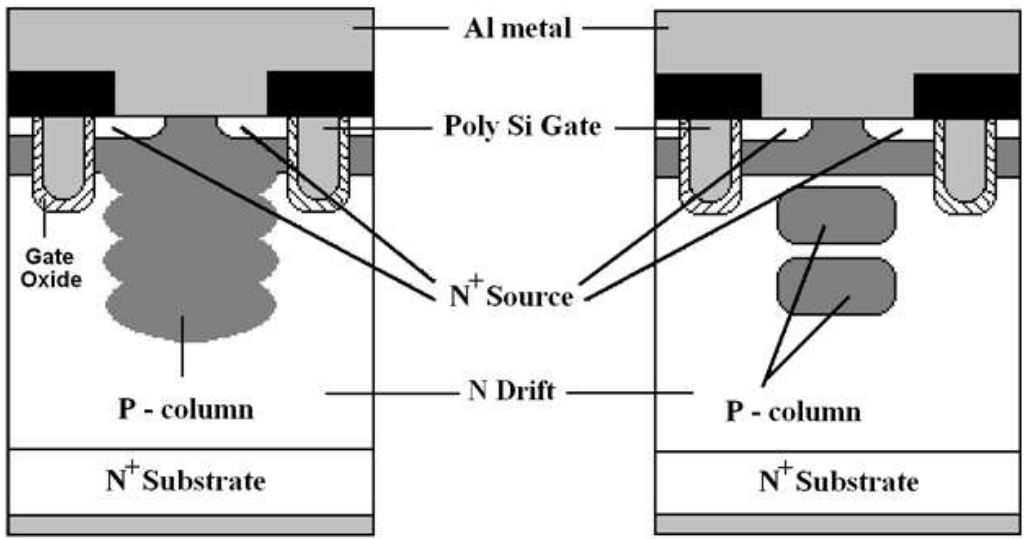

Fig. 4: A SJ trench MOSFET realized by implanting P-column base using multiple implants of boron in N-drift region, having (a) continuous P-base [66], (b) split P-base [67].

A new split gate, stepped oxide RESURF trench MOSFET is recently attracting a lot of research interest as it promises lower FOM. The structure has been examined experimentally [76] as well as through TCAD simulations [77] to have better switching characteristics due to lower gate charge. The use of double epi-layer drift region [78, 79] and asymmetric wing cell [80] are the other RESURF techniques proposed for improving the breakdown performance.

D.   *Use of Metal Substrate*

The major efforts made for reducing RON are usually based on the lateral scaling or pitch reduction. This approach is effective in reducing the channel resistance [81], but at the same time it causes gate charge to increase, which deteriorates the switching performance. Also, aggressively reducing the cell pitch causes the resistance contribution of silicon substrate to the total RON quite significant [82]. It has been shown that the 30 V n-channel trench gate MOSFETs, with a pitch of 4.0 μm, have 25% contribution of the specific RON from the silicon substrate having a resistivity of 3.5 mΩ.cm and thicknesses in the range of 200 μm [83]. Suppressing the substrate resistance contribution by using metal substrate, instead of silicon, may allow a significant reduction in the ON-resistance, without compromising gate charge. The incorporation of metal substrate is actually realized by transferring the silicon device layers to the metal substrate at the end of the device fabrication. This technology is known as silicon-on-metal (SOM) technology [84]. The trench MOSFET, fabricated using SOM technology on copper substrate, has demonstrated significant improvement in RON, energy efficiency and thermal conductance [84, 85]. The transient thermal resistance of the SOM device was also found to be much smaller than the silicon substrate devices. The increased biaxial compressive thermal stress in silicon, perpendicular to the channel was investigated by a 3D piezo-resistance model [85]. This stress results in an enhanced carrier mobility that in turn reduces the total device resistance. Additionally, an improved transient thermal conductance improves the ruggedness and reliability of the device.

## IV.     ADVANCEMENTS IN FABRICATION TECHNOLOGY

Being a device of commercial importance, the fabrication cost is the governing aspect of the trench gate MOSFET fabrication. Therefore, numerous attempts have been concentrated to cut down the fabrication cost while maintaining the high performance. A wide variety of methods for fabrication of trench devices have been inherited from the advanced semiconductor processing technology, but none of those has been found to be as superior in all respects as to displace all others. The fabrication usually involves a number of processing steps, such as, epilayer growth for drift region on a low resistivity silicon substrate and the growth of the body region over the substrate, trench creation by reactive ion beam etching, gate definition and contact-hole opening etc. All unit processes need special attention and online monitoring for consistently getting high yield. Out of various process steps, lithography and trench etching are the most important processes because a little deviation in these may largely degrade the performance and the overall yield. The impact of lithographic limitations on the device performance, as studied experimentally on 0.6 µm pitch devices [86, 87], indicates that scaling down the trench and contact width is achievable by DUV lithography.

To meet the requirement of low RON, a very high channel density is necessary that can be achieved by the self aligned fabrication process because otherwise the alignment tolerances between two lithography steps may become a bottleneck. Entire technological improvement efforts can be categorized in two: trench etching technology and self aligned process technology, because other process steps are not at all critical. The development of technologies and their issues are discussed in following subsection.

### A.     *Trench Etching Technology*

Opening high aspect ratio trenches reliably with controlled depth is the most important process step in realizing a high performance trench gate MOSFET. For reducing the RON, it is necessary to thin down the epitaxial layer and minimize the cell size which requires exact control of trench depth in order to maintain the desired drain-to-source breakdown voltage. Another important issue is the capacitive coupling between the gate and drain regions that governs the switching speed. The best switching performance can be achieved with a narrow trench that extends just beyond the P-N interface. The trench width is governed by the photolithographic capabilities and the compatibility with subsequent processing steps. The P-N junction depth can also be controlled accurately by ion implantation. Therefore, trench depth becomes the most critical parameter to control during the etching of silicon trenches. The smoothness of the trench sidewall that minimizes the gate leakage current, the sidewall profile that is required for its proper filling with poly Si, and trench bottom rounding to avoid the pre-matured device breakdown due to electric filed crowding at sharp corners are the other parameters which need to be considered for making a high performance, reliable trench device. The importance of trench etching process can be realized by the fact that the first successful realization of trench MOSFET, to an acceptable level, could be achieved only by an improved silicon trench processing technology [88, 89] and using a novel trench fabrication process only, the channel electron mobility approaching 87% of its bulk value could be obtained [90]. The control on trench etching process led to a narrow trench formation process and the double trench (one additional trench for contact) structure that enabled the 3rd generation, high performance trench gate MOSFET [91].

The trenches in a power MOSFET are usually etched using the $SF_6/O_2$ plasma. The balancing of bi-directional etching reactions between F atoms and the silicon substrate forms $Si_xF_y$ by-products. On the other hand, the oxygen in the plasma reacts with the etched silicon sidewalls to produce $Si_xO_y$ and thus provides the sidewall passivation. When an additional sidewall passivation is required, for example in case of deep trench opening, the $CF_4$ or $SiF_4$ is added into the process gas mix [92].

Since repeatability and yield have become the major governing forces for the power electronic devices, it is important to have a rugged and reliable fabrication process for the trench MOSFETs. The trench depth control during etching process, in a trench power MOSFET, is more difficult than that in many other dry etching applications because of the absence of stop layer where etching is to be terminated. Thus, the conventional etch stop technique of optical emission spectroscopy (OES) cannot be utilized for trench depth control and controlling that by adjusting the etching rate and time is also not a reliable method as it is subjected to many inconsistencies, affecting the repeatability and yield. Therefore, trench etching requires a sophisticated Reactive Ion Beam Etching (RIE) machine with in-situ depth monitoring system. To achieve the trench depth repeatability and yield up to the desired level, it is necessary to integrate a real time depth-monitoring and analysis tool embedded with the etching chamber itself. One of such techniques is the Interferometric End Point Detection (IEPD) which is also suited for high aspect ratio trenches. Implementation of this technique has been demonstrated successfully for trench device applications [93].

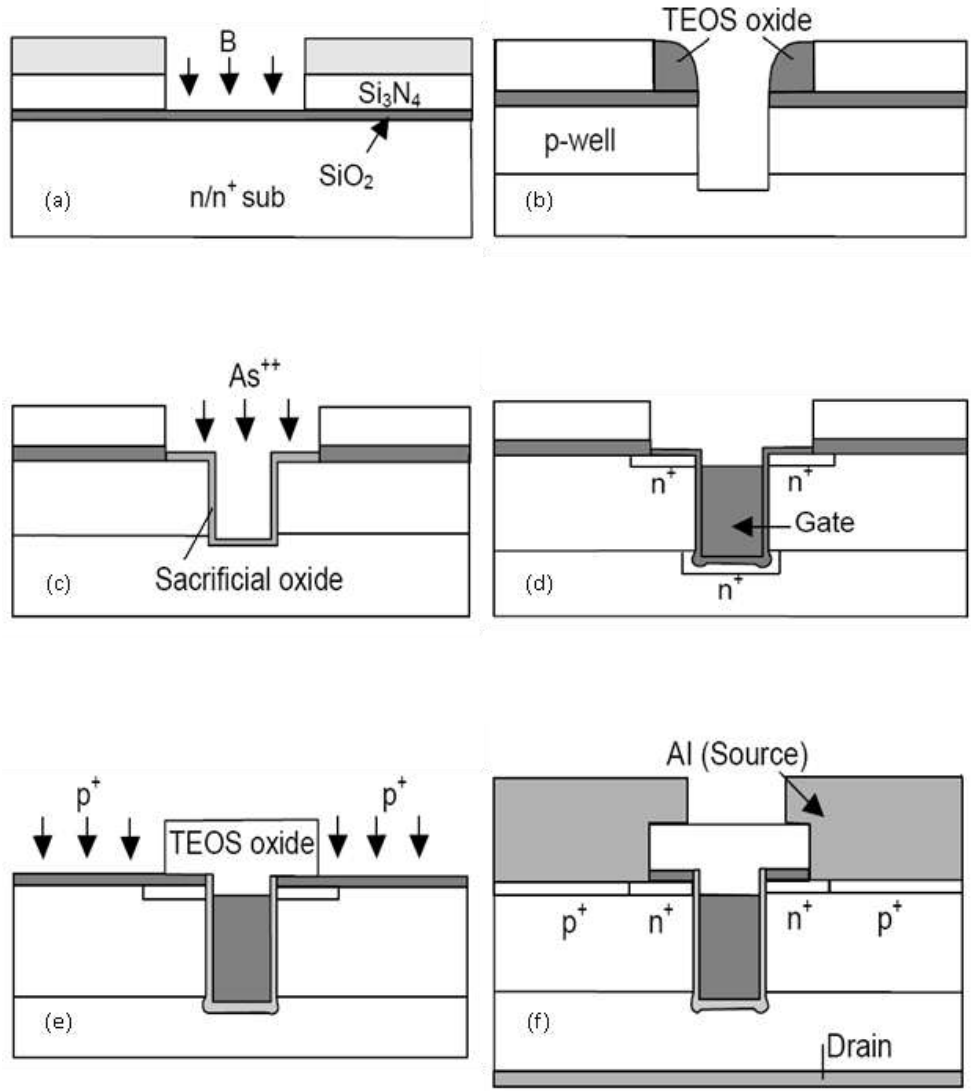

Fig. 5: Self aligned process sequence of fabricating trench MOSFET, using oxide spacer technique [102].

*B.    Self Aligned Fabrication Process*

The techniques of trench filling with Poly Si and etching it back along with selective etching of doped and undoped poly Si have been employed initially for realization of most of the processing steps, such as channel formation, gate definition, and contact-hole opening, through a single masking step [94, 95]. This has permitted a remarkable increase in packing density and therefore attracted a lot of attention and sincere efforts for further simplification and advancement in the self aligned process techniques. Shenai has suggested the use of poly Si pillar (formed while trench filling and removal of RIE mask) for spacer formation that gives self aligned contacts for source [39]. In another technique, local oxidation of poly Si pillar [96, 97] enabled self aligned contact holes for source/body connection. A well studied and implemented method of forming self aligned source was the use of oxide spacer [98-100] by deposition and RIE etching of TEOS oxide to expose surface areas of the N-epi where the trenches are to be formed. Then the exposed N-epi silicon and underlying P-body and N-drift region are etched away to define trenches [98, 101]. This approach was further modified by performing N+ ion implantation after removal of oxide spacer and also by making it more reliable by smoothing of trench corners using H2 annealing [37, 102-104]. A typical self aligned process sequence is shown in Fig. 5. Silicidation [105] and triple trench etching [106] were the other methods tried for making self aligned source/body contact.

A different approach of fabricating very high density trenches was demonstrated by J. Zeng, et. al. [107], wherein the poly Si gate is recessed into the trench, leaving a region to be filled by the BPSG the next step. The depth of the recess region determines the final thickness of BPSG, which is a design parameter for getting desired gate-source voltage rating. The BPSG layer is re-flown and is then etched back using the flat silicon surface as the end point. BPSG then serves as the isolation regions. In next step the BPSG isolation is planarized and the space occupied by BPSG at the surface is completely removed. This results in very closely spaced trenches. The finer process details are explained in [107]. A few more similar approaches of self aligned trench gate formation are discussed in [108, 109].

C.  *Other Technological Innovations*

The thicker gate oxide is preferred at the drain side to reduce the capacitive coupling of gate and drain regions that needs realization of larger bottom oxide thickness. The oxide bypassed structure utilizes the oxide thickness control in fabrication instead of the doping concentration control and has been shown to obtain RON versus VBD tradeoff limit beyond the conventional unipolar silicon limit [110, 111]. The structure also provides the option to utilize an additional bias voltage to compensate possible process variations in order to enhance the VBD. The structure, fabrication process and the laboratory results on the tunable oxide-bypassed trench gate power MOSFET have shown that this technology is feasible in making super junction MOSFET devices. Another innovative technique is the LOCOS process at the trench bottom that results in a W-shaped trench structure [42, 81, 112, 113] which promises 40% reduction in QG and about 58% improvement in RON.QG figure-of-merit [114, 115].

V.  **CHARACTERIZATION TECHNIQUES**

Characterization is an unavoidable and essential step in any semiconductor device manufacturing process because only using this step one can predict the final behavior of the device in its intended application before its actual deployment, saving a lot of time, money and efforts. Additionally, it provides feedbacks to the process engineers to avoid/minimize all the systematic and random errors in the device fabrication by fine tuning and controlling the process parameters to readjust the desired device parameters, if required, for yield improvement. Various standard techniques are used for material, process and device characterization [116] for this purpose.

Since in a trench MOSFET, majority of active parts are located deep in the silicon, the extraction of important device parameters like defect density in the epi-layer, doping-concentration variations in epi-layer (due to drive-in anneals and dopant out-diffusion), p-well doping concentration, and SiO2/Si interface quality is a challenging task. To overcome this problem, gate-controlled diode (GCD) current has been suggested to be used as the process control monitor [117]. Due to very high packing density, the effective transistor width becomes several meters. Therefore, W/L ratio becomes very large as the transistor channel length remains in the sub-micrometer range. The increased W/L ratio makes the total GCD current large enough for its reliable measurement. The GCD current has been shown to have signatures of the process conditions, such as silicon doping concentration in the epi-layer and effect of thermal cycles on it. The high-frequency and quasi-static CVs have also been used to extract the critical device parameters, like epi-doping-concentration, defect density in the epi, p-well doping concentration, and SiO2/Si interface quality [117].

The terminal current-voltage characteristics and gate charge transients are the usual methods to extract the performance parameters like ON-resistance, breakdown voltage, transconductance, gate charge and most importantly the FOM. Since, the total gate charge QG mainly depends on the gate to drain charge QGD, the figure-of-merit (FOM) is also defined as the product of RON and QGD that is commonly used to quantify the performance of the power MOSFETs for a specified off-state breakdown voltage. Additionally, measurement of carrier mobility in the channel, transient capability and energy handling capability are the other suggested and employed techniques to characterize the trench MOSFET.

### A.    *Gate Charge*

When the gate of a power MOSFET is connected to the supply voltage to turn it ON, first the capacitances associated with the gate are charged. The charging of these input capacitances decides the switching speed. The amount of charge supplied to the gate for charging these capacitances is defined as the total gate charge. Estimating the gate charge not only allows us to get an idea about the switching delays and losses, it also provides a comparison among the switching performances of two devices from different manufacturers. Thus, gate charge is a more useful parameter from the circuit design point of view as compared to the gate capacitance. Using experimental and analytical gate charge transient analysis [118, 119], it has been shown that QGD itself is composed of accumulation, depletion, and inversion charges. The inversion charge is located mainly underneath the trench bottom. The accumulation and depletion charge contribute each about 45% in conventional trench MOSFETs and can be reduced by using thick bottom oxide in a shallow trench device.

Another technique of analyzing gate charge utilizes a curve tracer. A curve tracer not only allows testing of the static parameters using DC characteristics, but can also show AC characteristics of the device. Due to the charging and discharging of capacitances of the device under AC signal, a phase difference exists between current and voltage. Therefore, the AC waveform of the power MOSFET generated on the curve tracer, serves to quantify its capacitance effect clearly and a qualitative idea about the various other capacitances may also be obtained, as discussed in [120].

### B.  *Carrier mobility extraction*

Extraction of carrier mobility in the channel is not straight forward. Various methods have been proposed to accurately estimate the value of carrier mobility in the channel of a trench MOSFET device. A method using fitting of experimental ON-resistance into an appropriate model has been proposed for extracting both the inversion and accumulation layer mobilities of electrons in n-channel trench MOSFET as a function of a wide range of effective electric field [121]. The split C-V method has been reported to be more accurate in evaluating the mobilities for the trench MOSFET which might have high interface states [122, 123].

### C.  *Energy capability estimation*

When a power MOSFET is switched ON or OFF, momentarily it is subjected to a very high current and high voltage simultaneously and therefore, may lead to excessive heat dissipation that may cause permanent device failure. The electrical or thermal instabilities within the device govern the Safe Operating Area (SOA) limits at the time of switching. Thus, designing of an optimum power device requires proper consideration for the SOA and its trade off with RON and VBD. The total energy that a device can safely handle is known as its energy capability (EC). The EC of a power MOSFET decides its energy switching capability and therefore is a very important parameter. The EC is measured by applying the controlled pulse train. The amplitude of the pulses and the duty cycle in increased continuously until the device fails. This measurement is applied in two different ways. In the first approach, an inductive load is switched OFF by the saw-tooth type power pulses, known as Unclamped Inductive Switching (UIS) [124]. In the second approach, the drain is clamped by a Zener diode to protect the device from avalanche operation, known as Clamped Inductive Switching (CIS) [125]. A theoretical and experimental comparison has been made between the two methods and it was found that in the CIS method, the measurement conditions are difficult to control independently and the results are dependent on the circuit parameters [126, 127] and therefore, a modified technique has been suggested wherein rectangular power pulses are applied to the device for excitation thereby directly forcing constant current or voltage [128]. This method provides the most straightforward control over the device excitation.

### D. Fabrication process control and monitoring

The process steps used to fabricate a trench gate power MOSFET are critical and need to be monitored and controlled for maintaining the desired performance and yield of the resulting device. The most critical step in the fabrication of trench gate power MOSFET is opening of trenches. Therefore careful monitoring of the trench depth, bottom, sidewalls and its subsequent filling with poly Si is very important for reliable device fabrication. The process characterization and monitoring need to be robust and should be applicable in the manufacturing environment. A conventional approach for process control in VLSI circuits is the use of MOS capacitor as a diagnostic tool. Similarly, the trench MOS capacitor may be analyzed using standard capacitance-voltage (CV) analysis and deep level transient spectroscopy (DLTS) to study dielectric/Si interface and bulk Si carrier traps in a trench gate MOSFET as reported in [129-131].

Model-Based Infrared Reflectometry (MBIR) is another useful technique for trench structure characterization [132]. It is used routinely for process control of deep trench DRAM devices and therefore can be employed in trench gate MOSFETs also for their process control. The technique is useful for measuring etch depth and monitoring the trench bottom at the time of trench opening. Furthermore, MBIR is also useful for detecting the voids in filled trenches and measuring poly Si fill depth after recess etch [133].

Various other techniques have also been employed for monitoring the trench gate MOSFET fabrication and control. The atomic force microscopy (AFM) has been shown to be effective for evaluating depth and resolving surface topological issues [134]. Use of secondary ion mass spectrometry (SIMS) has been demonstrated to be an effective approach for extraction of doping profiles in trenched regions within the MOSFET cells [135]. Scanning electron microscopy (SEM) has been suggested as an effective characterization tool for high-aspect-ratio three-dimensional structures with submicron dimensions [136] suitable for trenches. Angle-resolved X-ray photoelectron spectroscopy has been demonstrated to be very useful for monitoring of the surface composition of etched Si samples and depth profile in trenches [137]. The technique allows a full depth profile analysis of trenched structures by analyzing the exposed Si area, which is accessed in the trenches also, by using angled X-ray spectra.

## VI. TECHNIQUES OF RELIABILITY ESTIMATION AND ASSURANCE

Reliability is a very important issue for a device like trench power MOSFET that has commercial importance and need proper investigation and assurance for consistent yield. Usually, local defects due to electrical over stresses (EOS) occur during the backend processes like source metallization and the wire bonding [138] that can be avoided by proper care. However, there exist varieties of uncontrolled random mechanisms that may cause an early failure of a trench device. Some of the important concerns for early failure are the heavy ion radiation especially in the devices fabricated for space application, trench corners, oxide interface and hot electron degradation in addition to the usual process induced device failures, such as the poor interconnect of the poly gate and the metal, the bonding wire, and the etch process [118]. The aging failure is examined mainly through acoustic, electron and ion microscopy under extreme electro-thermal fatigue conditions. The de-lamination at the die attach is found to be one of the main aging modes related to the drastic increase in the drain resistance [138]. The avalanche behavior was investigated by measurements as well as electro-thermal simulations and found to be related to either of two mechanism, i.e., high energy consumption in the device and the high current being driven by it [139].

To make the trench MOSFET suitable for space applications, radiation hardness is another important issue as it may result in large shifts in I-V characteristics. Even a very small, micro-dose of the heavy ions may result in the parasitic transistor formation in the gate oxide and may have significant implications on the MOS devices for use in space. It is shown that they can lead to off-state leakage currents greater than 1.0 A [140, 141].

Other usual reliability concerns are related to the trench processing that primarily influence the device immunity to electrical stress, i.e., trench etching, trench cleaning and subsequent gate-oxide growth. Since the thermal-growth of gate oxide layer on the sidewall and base of a reactive-ion etched silicon surface is a three dimensional process, its interface with the Si at the bottom corners of the trench becomes the weakest region in the MOS structure and may cause serious reliability problems [39, 40, 129-132, 142-145]. Also, the thickness variations of the oxide layer at the sidewalls may result in the formation of voids while filling the trenches with poly Si. The oxide edge adjacent to the drain and the oxide/silicon interface are the most susceptible regions to damage [145].

The P-channel trench gate MOSFETs suffer from a peculiar reliability problem related to the Boron penetration from heavily boron-doped P+ poly-Si gate into the gate dielectric/Si-sidewall interface and into bulk Si [142], due to high diffusivity of Boron. The oxide-nitride complex gate structure is reported to overcome most of such reliability problems. It is shown that by optimizing this structure, high gate reliability same as that of a planar MOSFET can be obtained in trench gate MOSFETs [142]. The Boron doped (for body formation), n-channel trench MOSFETs have been examined using charge pumping technique [146] and it has been found that the hot electron effect is governed by the channel doping. The measurement of bulk current is a usual and conventional technique of estimating the hot carrier effect. However, recently the bulk current has been found to reach at its maximum at intermediate drain voltage [147]. This behavior has important consequences for the hot carrier reliability evaluation of the transistors and need a careful further investigation.

The hot free carriers, with sufficiently high energy, cross over the Si–SiO2 barrier and get trapped at the interface of the gate dielectric and the bulk. These trapped charges can induce large threshold voltage shifts due to the bias and temperature stress, resulting in the problems, known as negative and positive bias temperature instability (NTBI and PTBI). Even if the effect can partially be recovered or compensated externally during its standard operation, the large VTh shift can still affect the proper functionality of the device [148, 149] and therefore should not be under estimated. Moreover, the input and feedback capacitances are also affected by trapped charges. To analyze this, the device is stressed with gate voltage along with precise temperature steps and the gate-charge characteristics, before and after stress, is used to estimate shifts in capacitances (CGD and CGS). This indicates the amount of the degradation of physical properties under different stress time and stress temperature conditions [149].

## VII. CONCLUSIONS

We have presented a brief summary of the work done for the development of trench gate power MOSFETs. We have shown that starting from its evolution as the best power device in low to medium power applications; the trench gate power MOSFET has undergone several structural and technological changes, though its basic structure and operating concept remained unchanged. These modifications have made the present form of a trench MOSFET a reliable and better device. Since the ultimate goal of these improvements is to achieve the ideal specifications, the trench power MOSFET continues to evolve making it an even better device in future [150-153].

**REFERENCES**

[1]    Moores, H. T. Proc. Wescon IRE Convention Los Angeles, Aug. 1957, 63-73.

[2]    Nelson, H. Proc. IRE 1958, 46(6), 1062-1067.

[3]    Clark, M. A. Proc. IRE 1958, 46(11), 1185-1204.

[4]    Adler, M. S.; Owyang, K. W.; Baliga, B. J.; Kokosa, R. A. IEEE Trans. Electron Devices 1984, 31(11),1570-1591.

[5]    Hall, R. N. Proc. IRE 1952, 40(11), 1512-1518.

[6]    Teal, G. K. Proc. National IRE Conf. on Airborne Electronics Dayton, Ohio, May 10, 1954.

[7]    Heasell, E. I. IEEE Trans. Electron Devices 1978, 25(12), 1382-1388.

[8]    Bennett, W. P.; Kumbatovic R. A. IEEE Trans. Electron Devices 1981, 28(10), 1154-1162.

[9]     Sun, S. C.; Plummer, J. D. IEEE Trans. Electron Devices 1980, 27(2), 356-367.

[10]    Sigg, H. J.; Vendelin, G. D.; Cauge, T. P.; Kocsis, J. IEEE Trans. Electron Devices 1972, 19(1), 45-53.

[11]    Rodgers, T. J.; Meindl, J. D. IEEE Trans. Electron Devices 1973, 20(3), 226-232.

[12]    Holmes, F. E.; Salama, C. A. T. Electronics Letters 1973, 9(19), 457-458.

[13]    Rodgers, T. J.; Meindl, J. D. IEEE J. Solid State Circuits 1974, 9(5), 239-250.

[14]    Holmes, F. E.; Salama, C. A. T. Solid-State Electronics 1978, 17, 791-797.

[15]    Yoshida, I.; Kubo, M.; Ochi, S. IEEE J. Solid State Circuits 1976, 11(4), 472-477.

[16]    Farzan, B.; Salama, C. A. T. Solid-State Electronics 1976, 19, 297-306.

[17]    Bean, K. E. IEEE Trans. Electron Devices 1978, 25(10), 1185-1193.

[18]    Lisiak, K. P.; Berger, J. IEEE Trans. Electron Devices 1978, 25(10), 1229-1234.

[19]    Salama, C. A. T.; Oakes, J. G. IEEE Trans. Electron Devices 1978, 25(10), 1222-1228.

[20]    Bassous, E. IEEE Trans. Electron Devices 1978, 25(10), 1178-1185.

[21]    Koymen, H.; Smith, B. V.; Gazey, B. K. Electronics Letters 1979, 15(19), 601-602.

[22]    Ueda, D.; Takagi, H.; Kano, G. IEEE Trans. Electron Devices 1985, 32(1), 2-6.

[23]    Chang, H. R.; Temple, V. A. K.; Baliga, B. J. IEEE Trans. Electron Devices 1988, 35(12), 2459-2460.

[24]    Chang, H. R.; Holroyd, F. W. Solid Stare Electronics 1990, 33(3), 381-386.

[25]    Syau, T.; Venkatraman, P.; Baliga, B. J IEEE Trans. Electron Devices 1992, 39(11), 2672-2673.

[26]    Morancho, F.; Tranduc, H.; Rossel, P. Proc. 20th International Conference on Microelectronics (Miel'95) Serbia, Sep. 12-14, 1995, 692-694.

[27]    Morancho, F.; Tranduc, H.; Rossel, P. Proc. 21st International Conference on Microelectronics (MIEL-97) Yugoslavia, Sep.14-17, 1997, 395-398.

[28]    Wu, Y.; Tian, B.; Huang, H.; Hu, D.; Sin, J. K. O.; Kang, B. Proc. 20th International Symposium on Power Semiconductor Devices & IC's Oralando, FL, May 18-22, 2008, 127-130.

[29]    Ng, J. C. W.; Sin, J. K. O.; Guan, L. Proc. 20th International Symposium on Power Semiconductor Devices & IC's Oralando, FL, May 18-22, 2008, 91-93.

[30]    Ueda, D.; Takagi, H.; Kano, G. International Electron Device Meeting 1986, 32, 638-641.

[31]    Syau, T.; Venkatraman, P.; Baliga, B. J. Electronics Letters 1992, 28(9), 865-867.

[32]    Zeng, J.; Mawby, P. A.; Towers, M. S.; Board, K. Solid State Electronics 1995, 38(4), 821-828.

[33]     Zeng, J.; Mawby, P. A.; Towers, M. S.; Board, K. IEE Proc.-Circuits Devices Syst. 1996, 143(1).

[34]     Hattori, Y.; Suzuki, T.; Kodama, M.; Hayashii, E.; Uesugi, T. Proc. International Symposium on Power Semiconductor Devices & ICs Osaka, 2001, 427-430.

[35]     Narazaki, A.; Maruyama, J.; Kayumi, T.; Hamachi, H.; Moritani, J.; Hine, S. Proc. International Symposium on Power Semiconductor Devices Toulouse, France, May 22-25, 2000, 377-380.

[36]     Vankemmel, R. C.; Meyer, K. M. D. IEEE Trans. Electron Devices 1990, 37(1), 168-176, 1990.

[37]     Kim, J.; Roh, T. M.; Kim, S. G.; Lee, J. H.; Cho, K. I.; Kang, Y. I. IEEE Electron Device Letters 2001, 22(12), 594-596.

[38]     Shimizu, R.; Kuribayashi; H., Hiruta, R., Sudoh, K.; Iwasaki, H. Proc. 18th International Symposium on Power Semiconductor Devices & IC's Naples, Italy, Jun. 4-8, 2006.

[39]     Shenai, K. IEEE Trans. Electron Devices 1992, 39(6), 1435-1443.

[40]     Kim, S. G.; Roha, T. M.; Kim, J.; Park, I. Y.; Lee, J. W.; Koo, J. G.; Bae, I. H.; Cho, K. Journal of Crystal Growth 2003, 255, 123–129.

[41]     Rochefort, C.; Dalen, R. V. IEEE Electron Devices Lett. 2004, 25(2), 73-75.

[42]     Darvish M.; Yue, C.; Lui, K. H.; Giles, F.; Chan, B.; Chen, K.; Pattanayak, D.; Chen, Q.; Terrill, K.; Owyang, K. Proc. International Symposium on Power Semiconductor Devices & ICs Cambridge, UK, Apr. 14-17, 2003, 24-27.

[43]     Darvish, M. Proc. Bipolar/BiCMOS Circuits and Technology Meeting Sep. 28-30 2003, 15-21.

[44]     Brown, J.; Jaunay, S.; Darwish, M. Proc. PCIM Europe International Conference and Exhibition Nuremberg, Germany, May 20-22, 2003.

[45]     Darwish, M.; Yue, C.; Lui, K. H.; Giles, F.; Chan, B.; Chen, K. I.; Pattanayak, D.; Chen, Q.; Terrill, K.; Owyang, K. IEE Proc. Circuits Devices Syst. 2004, 151(3), 238-242.

[46]     Tsui, B. Y.; Wu, M. D.; Gan, T. C.; Chou, H. H.; Wu, Z. L.; Sune, C. T. Proc. International Symposium on Power Semiconductor Devices & ICs Kitakyushu, 2004, 213-216.

[47]     Narazaki, A.; Hisamoto, Y.; Tadokoro, C.; Takeda, M.; Hagino, H. Proc. International Symposium on Power Semiconductor Devices & ICs May 26-29, 1997, 285-288.

[48]     Juang, M. H.; Chen, W. T.; Ou-Yang, C. I.; Jang, S. L.; Lin, M. J.; Cheng, H. C. Solid-State Electronics 2004, 48, 1079–1085.

[49]     Juang, M. H.; Chueh, W. C.; Jang, S. L. Semicond. Sci. Technol. 2006, 21, 799–802.

[50]     Wang, Y.; Hu, H. F.; Cheng, C. Proc. IEEE International Conference on Electron Devices and Solid-State Circuits (EDSSC) Dec. 8-10, 2008.

[51]     Cezac, N.; Rossel, P.; Morancho, F.; Tranduc, H.; Peyre-Lavigne A.; Pages, I. Proc. 22nd International Conference on Microelectronics (MIEL 2000) NiS, Serbia, 2000, 637-640.

[52]     Cezac, N.; Morancho, F.; Rossel, P.; Tranduc, H.; Peyre-Lavigne, A. Proc. 12th International Symposium on Power Semiconductor Devices and ICs (ISPSD'2000) Toulouse, France, May 22-25, 2000, 69-72.

[53]     Morancho, F.; Cezac, N.; Galadi, A.; Zitouni, M.; Rossel P.; Peyre-Lavigne, A. Microelectronics Journal 2001, 32, 509-516.

[54]     Alves, S.; Morancho, F.; Reynes, J. M.; Lopes, B. Proc. 15th International Symposium on Power Semiconductor Devices and ICs (ISPSD'2003) Cambridge, UK, 2003, 308-311.

[55]     Takaya, H.; Miyagi, K.; Hamada, K.; Okura, Y.; Tokura, N.; Kuroyanagi, A. Proc. 17th International Symposium on Power Semiconductor Devices & IC's Santa Barbara, CA, May 23-26, 2005, 43-46.

[56]     Vaid, R.; Padha, N. Proc. 25th International Conference on Microelectronics (MIEL 2006) Belgrade, Serbia and Montenegro, May 14-17, 2006, 207-210.

[57]     Miyagi, K.; Takaya, H.; Saito, H.; Hamada, K. Proc. Power Conversion Conference (PCC) Nagoya, Apr. 2-5, 2007, 1011-1016.

[58]     Takaya, H.; Miyagi, K.; Hamada, K. Proc. 19th International Symposium on Power Semiconductor Devices & ICs Jeju, Korea, May 27-30, 2007, 197-200.

[59]     Takaya, H.; Miyagi, K.; Hamada, K. Proc. International Electron Devices Meeting, (IEDM '06) Dec. 11-13, 2006, 1-4.

[60]     Fujihara, T. Jpn. Journal Appl. Phys. 1997, 36, 6254-6262.

[61]     Chen, X. B.; Sin, J. K. O. IEEE Trans. Electron Devices 2001, 48(2), 344-348.

[62]     Chen, X. B.; Wang, X.; Sin, J. K. O. IEEE Trans. Electron Devices 2000, 47(6), 1280-1285.

[63]     Saito, W.; Omura, I.; Tokano, K.; Ogura, T.; Ohashi, H. IEEE Trans. Electron Devices, 2004, 51(5), 797-802.

[64]     Nitta, T.; Minato, T.; Yano, M.; Uenisi, A.; Harada, M.; Hine, S. Proc. International Symposium on Power Semiconductor Devices Toulouse, France, May 22-25, 2000, 77-80.

[65]     Hattori, Y.; Nakashima, K.; Kuwahara, M.; Yoshida, T.; Yamauchi, S.; Yamaguchi, H. Proc. International Symposium on Power Semiconductor Devices & ICs Kitakyushu, 2004, 189-192.

[66]     Ninomiya, H.; Miura, Y.; Kobayashi, K. Proc. International Symposium on Power Semiconductor Devices & ICs Kitakyushu, 2004, 177-180.

[67]     Miura, Y.; Ninomiya, H.; Kobayashi, K. Proc. 17th International Symposium on Power Semiconductor Devices & IC's Santa Barbara, CA, May 23-26, 2005, 39-42.

[68]     Kurosaki, T.; Shishido, H.; Kitada, M.; Oshima, K.; Kunori, S.; Sugai, A. Proc. International Symposium on Power Semiconductor Devices & ICs Cambridge, UK, Apr. 14-17, 2003, 211-214.

[69]     Chen, W.; Zhang, B.; Li, Z.; Xiang, J. Proc. International Conference on Communications, Circuits and Systems May 27-30, 2005, 1390-1394.

[70]     Sakakibara, J.; Suzuki, N.; Yamaguchi, H. Proc. 14th International Symposium on Power Semiconductor Devices and ICs Jun. 4-7, 2002, 233-236.

[71]     Yamaguchi, H.; Suzuki, N.; Sakakibara, J. Proc. International Symposium on Power Semiconductor Devices & ICs Cambridge, UK, Apr. 14-17, 2003, 316-319.

[72]     Sakakibara, J.; Urakami, Y.; Yamaguchi, H. Proc.  International Symposium on Power Semiconductor Devices & ICs Kitakyushu, 2004, 209-212.

[73]     Yamaguchi, H.; Urakami, Y.; Sakakibara, J. Proc. 18th International Symposium on Power Semiconductor Devices & IC's Naples, Italy, Jun. 4-8, 2006.

[74]     Yamauchi, S.; Shibata, T.; Nogami, S.; Yamaoka., T.; Hattori, Y.; Yamaguchi, H. Proc. 18th International Symposium on Power Semiconductor Devices & IC's Naples, Italy, Jun. 4-8, 2006.

[75]     Shibata, T.; Noda, Y.; Yamauchi, S.; Nogami, S.; Yamaoka, T.; Hattori, Y.; Yamaguchi, H. Proc. 19th International Symposium on Power Semiconductor Devices & ICs Jeju, Korea, May 27-30, 2007, 37-40.

[76]     Goarin, P.; Koops, G. E. J.; Van Dalen, R.; Camn, C. L.; Saby, S. Proc. International Symposium on Power Semiconductor Devices & ICs Jeju, Korea, May 27-30, 2007, 61 - 64.

[77]     Tong, C. F.; Cortes, I.; Mawby, P. A.; Covington, J. A.; Morancho, F. Proc. Spanish Conference on Electron Devices Santiago de Compostela, Spain, Feb. 1-13, 2009, 250 - 253.

[78]     Wang, Q.; Li, M.; Sharp, J.; Challa, A. IEEE Trans. Electron Devices 2007, 54(4), 833-839.

[79]     Li, M.; Crellin, A.; Ho, I.; Wang, Q. IEEE Trans. Electron Devices 2008, 55(7), 1749-1755, 2008.

[80]     Chien, F. T.; Liao, C. N.; Wang, C. L.; Chiu; H. C.; Tsai, Y. T. Electronics Letters 2008, 44(3).

[81]     Bulucea, C.; Rossen, R. Solid State Electronics 1991, 34(5), 493–507.

[82]     Baliga, B. J. Power Semiconductor Devices PSW Publishing: Boston, MA, 1996, 377–380.

[83]     Wang, Q.; Li, M.; Sim, G.; Ngo, A. Proc. Fairchild Semicond. Technol. Conf. San Diego, CA, 2004.

[84]     Wang, Q.; Li, M.; Sokolov, Y.; Black, A.; Yilmaz, H.; Mancelita, J. V.; Nanatad, R. IEEE Electron Devices Lett. 2008, 29(9), 1040-1042.

[85]     Wang, Q.; Ho, I.; Li, M. IEEE Electron Devices Lett. 2009, 30(1), 61-63.

[86]     Goarin, P.; Dalen, R. V.; Koops, G.; Cam, C. L. Proc. 36th European Solid-State Device Research Conference (ESSDERC 2006) Sep. 19-21, 2006, 274-277.

[87]     Goarin, P.; Dalen, R. V.; Koops, G.; Cam, C. L. Solid-State Electronics, 2007, 51, 1589–1595.

[88]     Tsui, B. Y.; Gan, T. C.; Wu, M. D.; Chou, H. H.; Wu, Z. L.; Sune, C. T. Proc. International Symposium on Power Semiconductor Devices & ICs, Kitakyushu, 2004, 205-208.

[89]     Shenai, K. IEEE Electron Device Letters 1991, 12(3), 108-110.

[90]     Shenai, K. Electronics Letters 1991, 27(9), 715-717.

[91]     Osawa, A.; Kanemaru, Y.; Matsuda, N.; Yoneda, T.; Matsuki, H.; Usui, Y.; Baba, Y. Proc. 11th International Symposium on Power Semiconductor Devices & ICs May 26-28, 1999, 209-212.

[92]     Bogart, K. H. A.; Klemens, F. P.; Malyshev, M. V.; Colonell, J. I.; Donnelly, V. M.; Lee, J. T. C.; Lane, J. M. J. Vac. Sci. Technol. A 2000, 18(1), 197-206.

[93]     Wei, W.; Zhongwen, L.; Wu, W.; Yungui1, G. Proc. 18th International Conference on Electronic Measurement and Instruments 2007, 251-254.

[94]     Ueda, D.; Takagi, H.; Kano, G. IEEE Trans. Electron Devices 1987, 34(4), 926-930.

[95]     Chang, H. R.; Black, R. D.; Temple, V. A. K.; Tantraporn, W.; Baliga, B. J. IEEE Trans. Electron Devices 1987, 34(11), 2329-2334.

[96]     Matsumoto, S.; Ohno, T.; Izumi, K. Electronics Letters 1991, 27(18), 1640-1642.

[97]     Matsumoto, S.; Ohno, T.; Ishii, H.; Yoshino, H. IEEE Trans. Electron Devices 1994, 41(5), 814-818.

[98]     Nam, K. S.; Lec, J. W.; Kim, S. G.; Roh, T. M.; Koo, J. G.; Cho, K. I. Electronics Letters 1999, 35(24), 2149-2150.

[99]     Matsuinoto, S.; Yoshino, H.; Ishii, H.; Ohno, T. Proc. 6th International Symposium on Power Semiconductor Devices & IC's Davos, Switzerland, May 31- Jun. 2, 1994, 365-369.

[100]   Narazaki, A.; Takano, K.; Oku, K.; Hamachi, H.; Minato, T. Proc. International Symposium on Power Semiconductor Devices & ICs Kitakyushu, 2004, 393-396.

[101]   Kim, J.; Roh, T. M.; Kim, S. G.; Song, Q. S.; Koo, J. G.; Nam, K. S.; Cho, K. I.; Ma, D. S. Proc. International Symposium on Power Semiconductor Devices Toulouse, France, May 22-25, 2000, 381-384.

[102]   Kim, J.; Roh, T. M.; Kim, S. G.; Park, I. Y.; Yang, Y. S.; Lee, D. W.; Koo, J. G.; Cho, K. I.; Kang, Y. I. ETRI Journal 2002, 24(5), 333-340.

[103]   Kim, J.; Roh, T. M.; Kim, S. G.; Park, I. Y.; Lee, B. IEEE Trans. Electron Devices 2003, 50(2), 378-383.

[104]   Baek, J.; Kim, J.; Kim, S. G.; Moona, J. K.; Lee, Y. H. Materials Science and Engineering B 2003, 97, 123-128.

[105]   Juang, M. H.; Sun, L. C.; Chen, W. T.; Ou-Yang, C. I. Solid-State Electronics 2001, 45, 169-172.

[106]   Park, I. Y.; Kim, S. G.; Koo, J. G.; Kim, J. Proc. International Symposium on Power Semiconductor Devices & ICs Cambridge, UK, Apr. 14-17, 2003, 169-172.

[107]   Zeng, J.; Dolny, G.; Kocon, C.; Stokes, R.; Kraft, N.; Brush, L.; Grebs, T.; Hao, J.; Ridley, R.; Benjamin, J.; Skurkey, L.; Benczkowski, S.; Semple, D.; Wodarczyk P.; Rexer, C. Proc. International Symposium on Power Semiconductor Devices & ICs Osaka, 2001, 147-150.

[108]   Kocon, C.; Challa, A.; Thorup, P. Proc. 18th International Symposium on Power Semiconductor Devices & IC's Naples, Italy, Jun. 4-8, 2006.

[109]   Kobayashi, K.; Kaneko, A.; Murase, Y.; Yamamoto, H. Proc. 19th International Symposium on Power Semiconductor Devices & ICs Jeju, Korea, May 27-30, 2007, 205-208.

[110]   Yang, X.; Liang, Y. C.; Samudra, G. S.; Liu, Y. Proc. 30th Annual Conference of the IEEE Industrial Electronics Society Busan, Korea,  Nov. 2 - 6, 2004, 729-733.

[111]   Aoki, T.; Tsuzuki, Y.; Miura, S.; Okabe, Y.; Suzuki, M.; Kuroyanagi, A. Proc.18th International Symposium on Power Semiconductor Devices & IC's Naples, Italy, Jun. 4-8, 2006.

[112]   Williams, R. K.; Grabowski, W.; Cowell, A.; Darwish M.; Berwick, J. Proc. International Symposium on Power Semiconductor Devices & IC's May 26-29, 1997, 193-196.

[113]   Kim, J.; Kim, S. G.; Roh, T. M.; Lee, B. Electronics Letters 2004, 40(11), 2149-2150.

[114]   Wang, H.; Trescases, O.; Xu, H. P. E.; Ng, W. T.; Fukumoto, K; Ishikawa, A. Furukawa, Y.; Imai, H.; Naito, T.; Sato, N.; Sakai, K.; Tamura, S.; Takasuka, K. Proc. 19th International Symposium on Power Semiconductor Devices & ICs Jeju, Korea, May 27-30, 2007, 181-183.

[115]   Wang, H.; Xu, H. P. E.; Ng, W. T.; Fukumoto, K.; Ishikawa, A.; Furukawa, Y.; Imai, H.; Naito, T.; Sato, N.; Sakai, K.; Tamura, S.; Takasuka, K. IEEE Electron Devices Lett. 2008, 29(11), 1239-1241.

[116]   Schroder, D. K. Semiconductor Material and Device Characterization 3rd ed., Wiley: NY, 2006.

[117]   Pan, J. IEEE Trans. Electron Devices 2009, 56(6), 1351-1354.

[118]   Hueting, R. J. E.; Hijzen, E. A.; Heringa, A.; Ludikhuize, A. W.; Zandt, M. A. A. IEEE Trans. Electron Devices 2004, 51(8), 1323-1330.

[119] Pan, S.; He, L.; Zhang, D. W.; Wang, L. K. Proc. 8th International Conference on Solid-State and Integrated Circuit Technology (ICSICT '06) Oct. 2006, 2132 - 2134.

[120] Pan, S.; He, L.; Wang, L. K.; Zhang, D. W. Proc. 8th International Conference on Solid-State and Integrated Circuit Technology (ICSICT '06) Oct. 2006, 254 - 256.

[121] Ng, J. C. W.; Sin, J. K. O. IEEE Trans. Electron Devices 2006, 53(8), 1914 - 1921.

[122] Yahata, A.; Inoue, T.; Ohashi, H. Applied Surface Science 1997, 117-118, 181-186.

[123] Heuvel, M. G. L. V. D.; Hueting, R. J. E.; Hijzen, E. A.; Zandt, M. A. A. Proc. International Symposium on Power Semiconductor Devices & ICs Cambridge UK, Apr. 14-17, 2003, 173-176.

[124] Phipps, P.; Gauen, K. Proc. 3rd Annual IEEE Applied Power Electronics Conference and Exposition (APEC'88) Feb. 1-5, 1988, 290-298.

[125] Farenc, D.; Charitat, G.; Dupuy, P.; Sicard, T.; Pages, I.; Rossel, P. Proc. 10th International Symposium on Power Semiconductor Devices & ICs Jun. 3-6, 1998, 359-362.

[126] Van den bosch, G.; Moens, P.; Gassot, P.; Wojciechowsk, D.; Groeseneken, G. Proc. 34th European Solid-State Device Research conference (ESSDERC) 2004, Sep. 21-23, 2004, 453-456.

[127] Passmore, L. J.; Sarpatwari, K.; Suliman, S. A.; Awadelkarim, O. O.; Ridley, R.; Dolny, G.; Michalowicz, J.; Wu, C. T. Thin Solid Films 2006, 504, 302-306.

[128] Merchant, S.; Baird, R.; Bennett, P.; Percy, P.; Dupuy, P.; Rossel, P. Proc. 10th International Symposium on Power Semiconductor Devices & ICs Jun. 3-6 1998, 317-320.

[129] Suliman, S. A.; Awadelkarim, O. O.; Fonash, S. J.; Ridley, R. S.; Dolny, G. M.; Hao, J.; Knoedler, C. M. Solid-State Electronics 2002, 46, 837-845.

[130] Suliman, S. A.; Venkataraman, B.; Wu, C. T.; Ridley, R. S.; Dolny, G. M.; Awadelkarim, O. O.; Fonash, S. J.; Ruzyllo, J. Solid-State Electronics 2003, 47, 899-905.

[131] Suliman, S. A.; Awadelkarim, O. O.; Ridley, R. S.; Dolny, G. M. Microelectronic Engineering 2004, 72, 247–252.

[132] Rosenthal, P. Proc. AIP Conference on Characterization and Metrology for ULSI Technology 2005, 620-624.

[133] Durán, C. A.; Maznev, A. A.; Merklin, G. T.; Mazurenko, A.; Gostein, M. Proc. IEEE/SEMI Advanced Semiconductor Manufacturing Conference 2007, 175-179.

[134] Ridley, R. S.; Strate, C.; Cumbo, J.; Grebs, T.; Gasser, C. Proc. IEEE/SEMI Advanced Semiconductor Manufacturing Conference 2002, 408-414.

[135] Zelsacher, R.; Wood, A. C. G.; Bacher, E.; Prax, E.; Sorschag, K.; Krumrey, J.; Baumgart, J. Microelectronics Reliability 2007, 47, 1585-1589.

[136] Harafuji, K.; Nomura, N. J. Appl. Phys. 1992, 72(7), 2541-2548.

[137]   Bogart K. H. A.; Donnelly, V. M. J. Appl. Phys. 2000, 87(12), 8351-8360.

[138]   Khong, B.; Legros, M.; Tounsi, P.; Dupuy, P.; Chauffleur, X.; Levade, C.; Vanderschaeve, G.; Scheid, E. Microelectronics Reliability 2007, 47, 1735–1740.

[139]   Pawel, I.; Siemieniec, R.; Rosch, M.; Hirler, F.; Herzer, R. IET Circuits Devices Syst. 2007, 1(5), 341–346.

[140]   Felix, A.; Shaneyfelt, M. R.; Schwank, J. R.; Dalton, S. M.; Dodd, P. E.; Witcher, J. B. IEEE Trans. Nuclear Science 2007, 54(6), 2181-2189.

[141]   Liu, S.; DiCienzo, C.; Bliss, M.; Zafrani, M.; Boden M.; Titus, J. L. IEEE Trans. Nuclear Science 2008, 55(6), 3231-3236.

[142]   Baba, Y.; Matuda, N.; Yawata, S.; Izumi, S.; Kawamura, N.; Kawakami, T. Proc. International Symposium on Power Semiconductor Devices & ICs May 26-29, 1998, 369-372.

[143]   Thapar, N.; Baliga, B. J. Solid-State Electronics 1997, 41(12), 1929-1936.

[144]   Dolny, G.; Gollagunta, N.; Suliman, S.; Trabzon, L.; Horn, M.; Awadelkarim, O. O.; Fonash, S. J.; Knoedler, C. M.; Hao, J.; Ridley, R.; Kocon, C.; Grebs, T.; Zeng, J. Proc. International Symposium on Power Semiconductor Devices & ICs Osaka, 2001, 431-434.

[145]   Suliman, S. A.; Gollagunta, N.; Trabzon, L.; Hao, J.; Ridley, R. S.; Knoedler, C. M.; Dolny, G. M.; Awadelkarim O. O.; Fonash, S. J. Semicond. Sci. Technol. 2001, 16, 447–454.

[146]   Suliman, S. A.; Awadelkarim, O. O.; Fonash, S. J.; Dolny, G. M.; Hao, J.; Ridley, R. S.; Knoedler, C. M. Solid-State Electronics 2001, 45, 655-661.

[147]   Moens, P.; Roig, J.; Desoete, B.; Bauwens, F.; Tack, M. IEEE Electron Devices Lett. 2008, 29(8), 909-912.

[148]   Aresu, S.; Kanert, W.; Pufall, R.; Goroll, M. Microelectronics Reliability 2007, 47, 1416–1418.

[149]   Alwan, M.; Beydoun, B.; Ketata, K.; Zoaeter, M. Microelectronics Journal 2007, 38, 727–734.

[150]   Saxena, R. S.; Kumar, M. J. IEEE Electron Device Lett. 2009, 30(9), 990-992, 2009.

[151]   Saxena, R. S.; Kumar, M. J. IEEE Trans. on Electron Devices 2009, 56(6), 1355-1359.

[152]   Saxena, R. S.; Kumar, M. J. IEEE Trans. on Electron Devices 2009, 56(3), 517-522.

[153]   Saxena, R. S.; Kumar, M. J. IEEE Trans. on Electron Devices 2008, 55(11), 3229-3304.